\documentstyle[tighten,floats,aps,epsf,preprint]{revtex}

\begin{document}
\title
{Exotic clusters in the excited states of $^{12}$Be, $^{14}$Be and $^{15}$B}

\author{Y. Kanada-En'yo}

\address{Institute of Particle and Nuclear Studies, \\
High Energy Accelerator Research Organization,\\
1-1 Oho, Tsukuba, Ibaraki 305-0801, Japan}

\maketitle
\begin{abstract}
The excited states of $^{12}$Be, $^{14}$Be and $^{15}$B were
studied by an antisymmetrized molecular dynamics method.
The theoretical results reproduced the energy levels of 
recently measured excited states 
of $^{12}$Be, and also predicted rotational bands 
with innovative clustering structures in $^{12}$Be, $^{14}$Be and 
$^{15}$B. Clustering states with
new exotic clusters ($^6$He, $^8$He and $^9$Li) were
theoretically suggested.
One new aspect in very neutron-rich nuclei is 
a 6-nucleon correlation among 4 
neutrons and 2 protons, which plays an important role in 
the formation of $^6$He clusters during clustering:
$^8$He+$^6$He of $^{14}$Be and $^9$Li+$^6$He of $^{15}$B.

\end{abstract}

Owing to progress of experimental techniques,
information concerning the excited states of light unstable nuclei 
has increased rapidly.
Clustering structures should
be one of the essential features in unstable nuclei
as well as in stable nuclei. However, the characteristics of 
clustering in unstable nuclei are still mysterious.
Do clustering structures often appear in the excited states 
of general light unstable nuclei ? 
Searching for exotic clustering structures and other new-type clusters than 
the well-known $\alpha$ cluster is the focus in studies of the 
excited states of light unstable nuclei. The clustering structures of 
unstable nuclei are one of the attractive and important subjects in  
experimental and theoretical research.

Theoretical studies on $^{10}$Be 
\cite{ENYOf,OGAWA,DOTE,ITAGAKI,ENYOg}
have suggested molecule-like structures in the $K=1^-$ band and
in the excited $K=0^+_2$ band.
Other candidates of clustering structures are 
the excited states of $^{12}$Be observed recently 
in the inelastic scattering reactions \cite{TANIHATA} and 
in break-up reactions into $^6$He+$^6$He and $^8$He+$^4$He
channels \cite{FREER,FREERb}.
However, fully microscopic calculations have not yet been done
for these states of $^{12}$Be.
Another interesting feature of $^{12}$Be is a
vanishing of the neutron magic number $N=8$ \cite{TSUZUKIa,IWASAKI}.
If the ground state is an intruder state, where are the normal states with
closed neutron-shell configurations?
It is important to solve the relation between the intruder ground 
state and the possible clustering states within one microscopic framework
in order to systematically investigate the structures of $^{12}$Be.
Our first aim is to study the 
ground and excited states of $^{12}$Be with a microscopic method. 
Furthermore, it is a challenge to theoretically search
light neutron-rich nuclei ($^{14}$Be and $^{15}$B)
for new clustering states with exotic clusters

We have applied a microscopic method of antisymmetrized molecular 
dynamics (AMD).
The AMD methods have already proved to be a powerful 
approach for structure studies
of general light nuclei \cite{ENYOg,ENYObc,ENYOe,ENYOsup}
because of the flexibility of the wave functions, which can express 
various kinds of structures from spherical shell-model-like structures
to developed clustering structures.
It is possible to systematically investigate 
the ground and excited states of light nuclei with a 
new version: a variation after spin-parity projections (VAP) of AMD,
which was proposed by the author \cite{ENYOe}.
As for theoretical predictions with this method,
the predicted $5^-$ and $6^+$ states of $^{10}$Be with 
developed clustering structures \cite{ENYOg} have actually been 
discovered in recent experiments \cite{FREERb}.

In the present work, we have studied the structures of the ground and 
excited states of $^{12}$Be and compared the theoretical results
with the experimental data. We theoretically searched $^{14}$Be and 
$^{15}$B for 
unknown excited states with new clustering structures. 
In the results for $^{14}$Be and $^{15}$B, the prediction 
of new rotational bands of exotic clustering structures are presented.
We discuss the mechanism of clustering 
development and the origin of 
the formation of $^6$He clusters.

An AMD wave function 
is a Slater determinant of Gaussian wave packets:

\begin{eqnarray}
&\Phi_{AMD}={1 \over \sqrt{A!}}
{\cal A}\{\varphi_1,\varphi_2,\cdots,\varphi_A\},\\
&\varphi_i=\phi_{{{\bf Z}}_i}\chi_i\tau_i :\left\lbrace
\begin{array}{l}
\phi_{{{\bf Z}}_i}({\bf r}_j) \propto
\exp\left 
[-\nu\biggl({\bf r}_j-{{\bf Z}_i \over \sqrt \nu}\biggr)^2\right],
\label{eqn:single}\\
\chi_{i}=
\left(\begin{array}{l}
{1\over 2}+\xi_{i}\\
{1\over 2}-\xi_{i}
\end{array}\right),
\end{array}\right.
\end{eqnarray}
where the centers of Gaussians ${\bf Z}_i$'s are complex variational
parameters. $\chi_i$ is an intrinsic spin function parameterized by
$\xi_{i}$, which is also a variational parameter specifying the direction 
of the intrinsic spin of the $i$th nucleon.
 $\tau_i$ is an isospin
function which was fixed to be up (proton) or down (neutron)
 in the present calculations.
We varied the parameters ${\bf Z}_i$ and $\xi_{i}$($i=1\sim A$) to
minimize the energy expectation value for the 
spin-parity eigenstate (VAP calculations), 
\begin{equation}
{\langle P^J_{MK'}\Phi^\pm_{AMD}|H|P^J_{MK'}\Phi^\pm_{AMD}\rangle \over
\langle P^J_{MK'}\Phi^\pm_{AMD} |P^J_{MK'}\Phi^\pm_{AMD}\rangle },
\end{equation}
where the operator of total-angular-momentum projection ($P^J_{MK'}$) is 
$\int d\Omega D^{J*}_{MK'}(\Omega)R(\Omega)$.
The integration for the Euler angle $\Omega$ was calculated numerically.
We adopted a frictional cooling method 
\cite{ENYObc} to obtain the minimum energy states. 
Here, we represent the intrinsic AMD wave function 
as $\Phi^{J\pm}_0$, that was obtained by a VAP calculation
for the lowest state with a given spin parity of $J^\pm$.
Higher excited states were constructed by varying the energy of the
orthogonal components to the lower states by superposing wave functions. 
More details of the AMD method for the excited states 
with the variation after spin-parity projection are described in Refs 
\cite{ENYOg,ENYOe}.
By performing VAP calculations for the lowest and 
the higher excited states with various sets of total spin and parity
$\{J\pm\}$,
we obtained many AMD wave functions $\{\Phi_1,\cdots,\Phi_m\}$, 
each of which approximately represents the intrinsic state of the 
corresponding $J^\pm_n$ state.
The final results were attained by diagonalizing 
a Hamiltonian matrix,
$\langle P^{J\pm}_{MK'}\Phi_i|H|
P^{J\pm}_{MK''}\Phi_j\rangle$ ($i,j=1\sim m$).

We applied the AMD method for the excited states of $^{12}$Be, $^{14}$Be
and $^{15}$B. The adopted interactions in the present work were 
the central force of the modified Volkov No.1 with case 3 \cite{TOHSAKI},
the spin-orbit force of G3RS \cite{LS} and the Coulomb force.
The Majorana parameter used here was $m=0.65$, 
and the strength of G3RS force was chosen to be  
$u_1=-u_2=3700$ MeV. We choose an optimum width parameter 
($\nu$) for the Gaussians of the single-particle wave functions 
of each nucleus. 

The theoretical results of the binding energies of 
$^{12}$Be, $^{14}$Be and $^{15}$B are 
61.9 MeV, 59.7 MeV and 73.1 MeV, which underestimate the experimental
values: 68.65 MeV, 69.77 MeV, 88.19 MeV, respectively.
We find that the smaller values of the binding energies in the present results
than the experimental data can be improved by changing the Majorana 
parameter of the interaction small to be $m=0.61$. 
With this parameter, the binding energy of 
$^{14}$Be is 67.7 MeV. 
In spite of the smaller values of binding energies, 
we adopt the parameter $m=0.65$ in the present work, because this parameter
gives a good reproduction of parity inversion of $^{11}$Be, which 
should be important to well describe the ground-state properties of 
neighboring nucleus: $^{12}$Be.
We have checked that the change of $m$ parameter has no 
significant effect on 
the excitation energy $E(0^+_2)$ of $^{14}$Be, at least.
Namely, the calculated values of $E(0^+_2)$ in the cases $m=0.61$ and 
$m=0.65$ are 4.2 MeV and 4.4 MeV, respectively.

\begin{figure}
\noindent
\epsfxsize=0.45\textwidth
\centerline{\epsffile{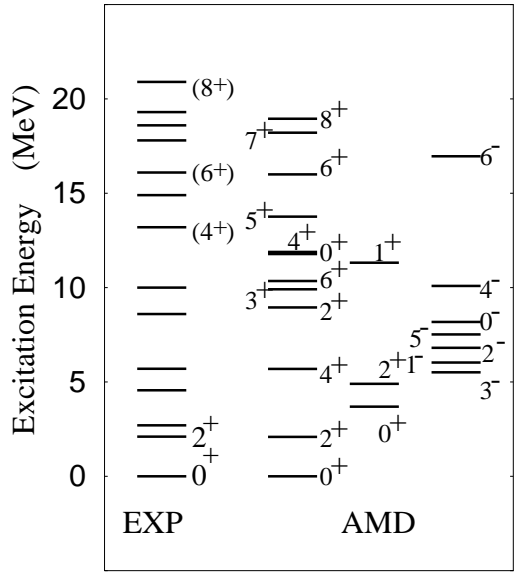}}
\caption{\label{fig:be12spe}
Excitation energies of the levels in $^{12}$Be. 
The theoretical results were calculated by VAP calculations based on AMD.
The experimental data are from the Table of Isotopes and Refs 
\protect{\cite{TANIHATA,FREER,IWASAKI}}.
}
\end{figure}

In the calculated results for $^{12}$Be with AMD, 
many excited states appear in the low-energy region.
The energy levels of $^{12}$Be are presented in
Fig. \ref{fig:be12spe}. 
The theoretical levels of the $4^+_2$, $6^+_2$ and $8^+_1$ states
correspond well to the recently 
observed excited states \cite{FREER}.
By analyzing the intrinsic AMD wave functions, we can classify 
the excited states into rotational bands, such as 
$K^\pi=0^+_1,0^+_2,0^+_3,1^-_1$. It is surprising that  
the newly observed levels \cite{FREER} at an energy region above 10 MeV 
belong to the third rotational band, $K^\pm=0^+_3$, which is a 
$2\hbar\omega$ excited state with a well-developed clustering structure.
On the other hand, there are many low-lying positive-parity states 
which belong to the lower
rotational bands, $K^\pi=0^+_1$ and $K^\pi=0^+_2$.
Even though $^{12}$Be has a neutron magic number of 8,
the intrinsic state of the ground $0^+$ state 
is not an ordinary state with a closed neutron $p$-shell, 
but a prolately deformed state with a developed 
clustering structure. The deformed ground state is dominated by another
$2p-2h$ configuration than that in the $0^+_3$ state.
As a result, the ground $K^\pi=0^+$ band starts
from the ground $0^+$ state and reaches the band terminal at
the $8^+_1$ state.
The rotational bands, $K^\pm=0^+_3$ and $K^\pm=0^+_1$,
share the same band terminal state because
the terminal $J^\pm=8^+_1$ state is the highest spin state in the 
$2\hbar\omega$ configurations.
On the other hand, the main components of the $0^+_2$ and $2^+$  states are 
the $0\hbar\omega$ configuration with a closed neutron $p$-shell,
and construct the second $K^\pi=0^+_2$ band.

For experimental evidence, the strength of the $\beta$ decay 
from $^{12}$Be(0$^+_1$) to $^{12}$B($1^+$) is helpful to estimate the
breaking of the neutron $p$-shell \cite{TSUZUKIa}.
As expected, the theoretical value of $B(GT)= 0.8$ 
is sufficiently small as the experimental data, $B(GT)=0.59$, because 
the component of the $2\hbar\omega$ configurations
in the parent $^{12}$Be(0$^+_1$) makes the transition matrix 
of the Gamov-Teller operator to be small. 
Such a result is consistent with pioneer work \cite{TSUZUKIa,ITAGAKIa},
where their treatments were not fully microscopic for 
the freedom of all nucleons.
The calculated value of $B(E1)$ from $1^-_1$ to $0^+_1$ is 0.02 $e^2$fm$^2$,
which agrees reasonably with the experimental large value, 
$B(E1)$ = 0.05 $e^2$fm$^2$ \cite{IWASAKI}. 
As for the $E2$ transition strength, 
the predicted values in the present results are 
$B(E2;2^+_1\rightarrow 0^+_1)$ = 14 $e^2$fm$^4$ and 
$B(E2;2^+_2\rightarrow 0^+_2)$ = 8 $e^2$fm$^4$, which indicate 
a larger value of the $E2$ transition strength in the deformed $K=0^+_1$ band
than that in the $K=0^+_2$ band.

It should be pointed out that this is the first theoretical work 
which systematically reproduces the energy levels of all  
spin-assigned states discovered recently, except for the $1^-$ state.
Although the excitation energy of the $1^-$ state is overestimated by 
theory, we regard the theoretical $1^-_1$ state as being 
the measured $1^-$ state at 2.68 MeV because of the large $B(E1)$ value.
A theoretical discovery in the present calculations is
the $K^\pi=0^+_3$ rotational band with
the developed $^{6}$He+$^6$He clustering structure, which constructs
the $0^+_3$, $2^+_3$, $4^+_2$, $6^+_2$ and $8^+_1$ states. 
The developed clustering in the $0^+_3$ state 
gradually weakens  with increasing total-angular
momenta ($J$), and finally it changes to a spin-aligned state at $J^\pm=8^+_1$.
Many excited states in the rotational bands $K^\pi=0^+_1,0^+_2$ and $1^-$
are predicted to exist in the low-energy region with an excitation energy 
below 10 MeV.

By analyzing the single-particle wave functions in the 
intrinsic state, the deformation mechanism of 
the ground state can be understood by the idea of molecular orbits
surrounding 2$\alpha$ cores. Although the molecular orbits in 
Be isotopes have been studied theoretically 
\cite{ENYOf,DOTE,ITAGAKI,OERTZEN},
one of the important points of the present results is that 
the development of a $2\alpha+4n$ structure in the ground state of $^{12}$Be 
can be confirmed without assuming the existence of clusters.
On the other hand, the third $0^+$ state seems to be
a $^6$He+$^6$He clustering structure instead of the $\alpha$ and surrounding
neutrons in molecular orbits, 
because the distance between two clusters is too large.
In this case, an $\alpha$ cluster goes outward
far enough to form a $^6$He cluster with 2 valence neutrons.

\begin{figure}
\noindent
\epsfxsize=0.45\textwidth
\centerline{\epsffile{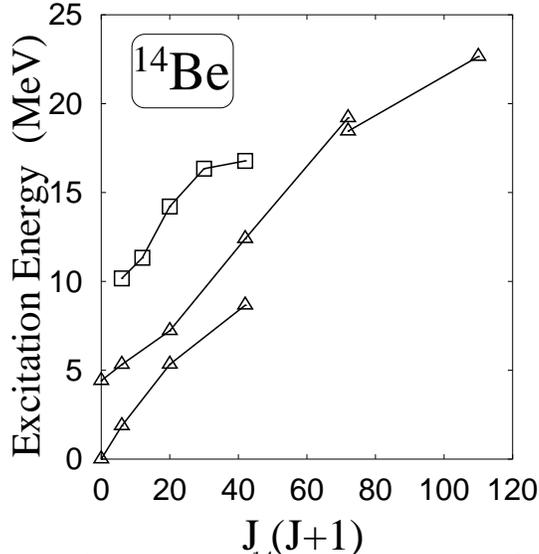}}
\caption{\label{fig:be14spe}
Excitation energies of the levels in $^{14}$Be
calculated by VAP calculations based on AMD.
The excitation energies are plotted as a function of $J(J+1)$,
where $J$ is the total spin of the state.
}
\end{figure}

We next present the results for $^{14}$Be.
Figure \ref{fig:be14spe} shows the energy levels of positive-parity 
states of $^{14}$Be obtained by VAP calculations.
Although few excited states are experimentally known, 
many excited states are predicted based on the theoretical results.
There are a few candidates of levels 
above the threshold energy (9.1 MeV excitation) for 
separation into $^8$He+$^6$He in break-up experiments \cite{SAITO}.

\begin{figure}
\noindent
\epsfxsize=0.45\textwidth
\centerline{\epsffile{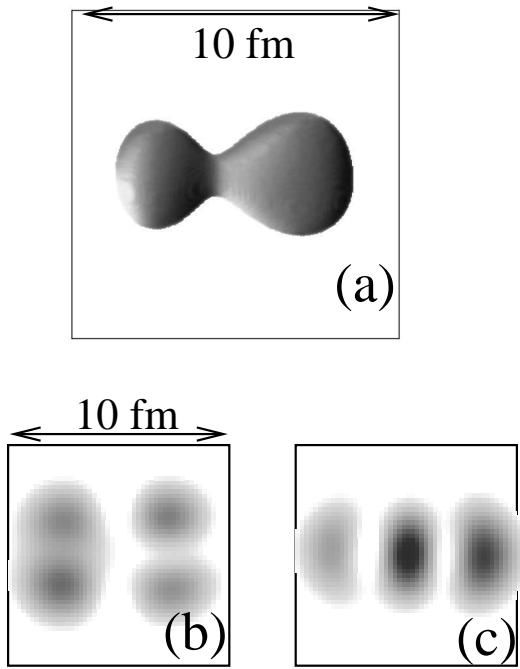}}
\caption{\label{fig:be14}
Intrinsic structure of $0^+_2$ of $^{14}$Be.
Figure (a) shows a surface cut of the matter density $\rho \ge 0.16$ 
nucleons/fm$^3$. The density was calculated for the
intrinsic AMD wave function before a spin-parity projection.
The density distributions for the single-particle orbits of 
the valence neutrons are presented in figures (b) and (c).
}
\end{figure}

 The excited states in the region $J\le 6$ 
are classified into 3 rotational bands:
$K=0^+_1$, $K=0^+_2$ and $K=2^+$.
In most of the states in $^{14}$Be, 2-$\alpha$ cores are formed in the 
calculated results, although no 
clusters are assumed in the model. 
Because of the 2-$\alpha$ cores, the $0^+_1$ and $0^+_2$ states 
have prolate deformations with deformation parameters of
$\beta=0.49$ and $\beta=0.64$, respectively. The latter one with a large 
deformation constructs the $K^\pi=0^+_2$ rotational band, which reaches
the $8^+_2$ state. 
By analyzing the single-particle wave functions of the intrinsic states,
it is found that the states in the $K=0^+_1$ and $K=2^+$ bands 
are dominated by the normal $0\hbar\omega$ configurations,
 while the largely deformed states in $K^\pi=0^+_2$ mainly contain
$4p$-$2h$ states with $2\hbar\omega$ configurations.
The unique result of $^{14}$Be is 
innovative clustering structures of the excited states 
in this second $K^\pi=0^+_2$ band. 
The intrinsic states of $K^\pi=0^+_2$ band
 have spatially developed $^8$He+$^6$He clustering 
state, where both clusters are,  themselves, very neutron-rich nuclei.
In Figure \ref{fig:be14}(a) for the contour surface for the matter density,
$\rho\ge$ 0.16 nucleons/fm$^3$, of the intrinsic state of the 
$0^+_2$ state, the $^8$He($^6$He) cluster can be clearly recognized 
in the right(left)-handed side.

In order to understand the origin of
$^8$He+$^6$He clustering formation, it is valuable to inspect the
single-particle behavior of the valence neutrons.
As mentioned above, the $^8$He+$^6$He clustering state
in the $K^\pi=0^+_2$ band comes from the $4p$-$2h$ configurations.
According to analyses of the single-particle energies and wave functions,
the highest 4 neutron orbits correspond to two 
kinds of spatial orbits, which are associated with the $sd$-mixing orbits 
in the deformed system.
Roughly speaking, each of two kinds of spatial orbits is occupied
by a spin-up neutron and a spin-down neutron.
Figures \ref{fig:be14}(b) and \ref{fig:be14}(c) present the higher $sd$-like 
orbit and the lower $sd$-like orbit, respectively. They contain 
more than 80\% components of the positive parity states.
If we call the longitudinal 
direction of the prolate deformation as the $z$-axis,
the highest nucleon orbit seen in Fig. \ref{fig:be14}(b) 
originates from the $sd$-orbit with a form of $yz \exp [-\nu r^2]$.
On the other hand, the spatial orbit for the lower one
is similar to the orbit of
$z^2 \exp[-\nu r^2]$, which deforms along the $z$-axis, as shown in 
Fig.\ref{fig:be14}(c). The original $sd$-orbits are modified because of 
the deformation in $^{14}$Be. 
The deformed neutron orbits are stabilized by 2 stretched pairs of 
protons, because the orbits gain their kinetic energies with respect to the
nodes along the $z$-axis.
From the viewpoint of single-particle orbits, 
the $^6$He cluster in the $^6$He+$^8$He clustering system consists of 
4 neutrons in the $sd$-orbits and 2 protons in the $p$-orbits.
In other words, the formation of $^6$He cluster  
originates from the correlation of 
4 neutrons in the $sd$-orbits and 2 protons in the $p$-orbits.
Once a $^6$He cluster takes shape, the cluster spatially develops out of the
core-cluster $^8$He keeping the stability of the modified 
$sd$-mixing orbits in the deformed system.
Unique points of this idea are as follows.
Firstly, correlations of more than 4 nucleons have not been found.
The $^6$He cluster is related to 6-nucleon
correlations (4 neutrons with 2 protons).
Second, the nucleons, which compose a cluster out of the core nucleus, 
rise over 2 shells ($sd$-shell and $p$-shell in this case).
This is unusual in light stable nuclei.
The third point is that the spatially clustering development
is associated with the stability of the single-particle orbits of the system.
We suggest one of the novel feature
that 4 neutrons with 2 protons correlate to form a developed
$^6$He cluster in neutron-rich nuclei. 

In the results concerning the excited states of $^{15}$B, 
the situation is similar to that of $^{14}$Be. We find an exotic
clustering structure with $^9$Li+$^6$He in the second 
$K^\pi=3/2^-_2$ band, which starts from the $J^\pm=3/2^-_2$ 
state at about 7 MeV higher than the ground state.
The intrinsic structure of the $K^\pi=3/2^-_2$ band consists
 of two exotic clusters:$^9$Li and
$^6$He clusters, which develop as remarkably as those in the 
$K^\pi=0^+_2$ band of $^{14}$Be.
These excited states with $^9$Li+$^6$He clustering
are based on $2\hbar\omega$ configurations.
The spatial parts of 4 valence neutrons in $sd$-like orbits 
are very similar to those of $^{14}$Be presented 
in Figs. \ref{fig:be14}(b) and 
\ref{fig:be14}(c). 
It is an interesting problem whether or not the $^6$He-cluster
may appear in the excited states of other neutron-rich nuclei 
than $^{14}$Be and $^{15}$B.

Although particle decay widths are important for the stability of 
the excited states,
the widths were not considered in the present calculations,
which were performed within a bound state approximation.
The widths for the particle decays should be
carefully discussed using other frameworks, such as a method with
a reduced width amplitude, or a complex scaling method
beyond the present AMD framework.
We just mention the partial widths of the states into He channels in the
$K^\pi=0^+_2$ of $^{14}$Be.
When the excitation energies are above the threshold energy, 12.2 MeV,
the excited states in $K^\pi=0^+_2$ of $^{14}$Be may decay into
$^8$He+$^6$He channel.
By assuming $^8$He and $^6$He clusters in SU(3) limits,
we estimated the partial decay width of $6^+_2$ and $8^+_2$ 
into $^8$He(0$^+$)+$^6$He(0$^+$) channel to be 6 keV and 50 keV, respectively,
by calculating the reduced width amplitude.
For the total width, it is required to carefully study the excitation 
energies and other decay channels, such as neutron decays 
and excited He cluster decays.

In summary, we studied the structure of the excited states of $^{12}$Be,
$^{14}$Be and $^{15}$B based on the framework of the AMD method. 
This is the first microscopic calculation to systematically reproduce the 
energy levels of all the spin-assigned states in $^{12}$Be, 
except for the $1^-$ state. 
One of the discoveries of the present results is the third $K^\pi=0^+$ band
with $^6$He+$^6$He clusters, which corresponds well to recently 
observed states in the break-up reactions.
In $^{14}$Be and $^{15}$Be, we predicted low-lying excited states with 
innovative clustering structures. Namely, new types of clusters, $^8$He+$^6$He
 and $^{9}$Li+$^6$He, were found
in the excited states of $^{14}$Be and $^{15}$Be, respectively.
In these well-developed clustering structures with 2$\hbar\omega$ excited
configurations, the mechanism of the $^6$He cluster development was 
discussed in relation to the single-particle picture.
It was found that the 6-nucleon correlation among 4 neutrons and 2 protons 
plays an important role in the formation of $^6$He cluster.
The results suggest an important
feature, that an exotic clustering structure may very often exist 
in the excited states of neutron-rich nuclei.

The author would like to thank Prof. H. Horiuchi for many discussions.
She is also thankful to Dr. N. Itagaki and Prof.
W. Von Oertzen for helpful discussions 
and comments. Valuable comments of Prof. S. Shimoura and A. Saito
are also acknowledged.
The computational calculations of this work are supported by 
RCNP in Osaka University, YITP in Kyoto University and
the supercomputer in KEK.

\end{document}